\documentclass[twocolumn,showpacs,preprintnumbers,amsmath,amssymb]{revtex4}

\usepackage{graphicx}
\usepackage{dcolumn}
\usepackage{bm}

\begin{document}

\preprint{APS/123-QED}


\title{Microwave  Photoconductivity
in Two-Dimensional Electron Systems due to Photon-Assisted
Interaction of Electrons with Leaky Interface Phonons} 
\author{ 
V.~Ryzhii }
\email{v-ryzhii@u-aizu.ac.jp}
\affiliation{Computer Solid State Physics Laboratory, University of Aizu,
Aizu-Wakamatsu 965-8580, Japan}


\date{\today}

\begin{abstract}
We calculate the contribution
of the photon-assisted interaction of electrons with
leaky interface phonons to the dissipative dc photoconductivity
of a two-dimensional electron system in a magnetic field.
The calculated photoconductivity as a function
of the frequency of microwave radiation and the magnetic
field exhibits pronounced oscillations.
The obtained oscillation structure is different from that
in the case of photon-assisted interaction with impurities.
We demonstrate that at a sufficiently strong microwave radiation
in the certain ranges of its frequency (or in certain ranges
of the magnetic field)
this mechanism can result in the absolute negative conductivity.
\end{abstract}



\maketitle

A substantial interest in the transport phenomena
in  a two-dimensional electron system (2DES) subjected to a magnetic field
 has been revived
after experimental observations by Mani {\it et al.}~\cite{1}, 
Zudov {\it et al}.~\cite{2,3} (see also Ref.~\cite{4}) of
vanishing electrical resistance 
caused  by microwave radiation.
The occurrence of this effect is primarily attributed to
the realization of the absolute negative conductivity (ANC) when
the dissipative dc conductivity $\sigma < 0$ in certain
ranges of the microwave frequencies and the magnetic and
electric fields~\cite{5,6,7,8}.
Mechanisms of ANC in a 2DES subjected to a magnetic field
and irradiated with microwaves have been studied theoretically
over many years~\cite{9,10,11,12,13,14}.
The mechanisms in question are associated with
the photon-assisted electron-impurity and  electron-phonon
interactions. As shown~\cite{15},
the interaction of electrons with leaky interface phonons
can  essentially govern the transport in a 2DES.
The two dimensional character of the spectrum of such phonons
substantially affects the scattering selection rules. As a result,
the contributions to dissipative dc conductivity
 of the electron
scattering processes with  2D and 
3D acoustic phonons can be markedly different.
The dissipative dc conductivity of a 2DES in the magnetic field
in ``dark''
conditions (without irradiation)
determined by the electron interaction with  2D acoustic phonons
was calculated in Refs.~\cite{15,16}. The effect of   
the electron interaction with  3D-acoustic phonons 
on the dc conductivity and the dc microwave photoconductivity
was considered recently~\cite{14}.

In this paper, we calculate the dissipative dc photoconductivity of a 2DES 
in the magnetic field irradiated with microwaves considering
the photon-assisted interaction of electrons with leaky interface phonons.
In particular, it is  shown that this mechanism can 
lead to   ANC.

The dissipative transport
in the situation under consideration is associated with
the shifts (hops) of the electron  Larmor orbit
centers in the direction of the net dc
electric field ${\bf E} = (E, 0,0)$ and in the opposite
direction caused by  the electron scattering.
The length of such a shift equals $\delta \xi = L^2q_y$,
where $L = (c\hbar/eH)^{1/2}$
is the quantum Larmor radius,  
$q_y$ is the variation the electron momentum
component perpendicular to the direction of the electric field,
$e = |e|$ is the electron charge, $\hbar$ is the Planck constant,
${\bf}H = (0, 0, H)$ 
is the magnetic field (directed perpendicular to the 2DES plane), and 
$c$ is the velocity of light. Taking this into account,
we start from the following sufficiently general expression
for the dissipative dc current in a 2DES in the presence of 
microwave radiation: 
$$
j_{ph}(E) = \frac{2\pi e}{\hbar L} \sum_{N,N^{\prime} }
f_N(1 - f_{N^{\prime}})
$$
$$
\times\int dq_xdq_y\,
q_yI_{\Omega}(q_x,q_y)|V(q_{\perp})|^2|Q_{N,N^{\prime}}(L^2q^2_{\perp}/2)|^2
$$
$$
\times\{{\cal N}_{q_{\perp}}
\delta[\hbar\Omega + (N - N^{\prime})\hbar\Omega_c + \hbar sq_{\perp} 
+ eEL^2q_y]
$$
\begin{equation}\label{eq1}
+ ({\cal N}_{q_{\perp}} + 1)
\delta[\hbar\Omega + (N - N^{\prime})\hbar\Omega_c - \hbar sq_{\perp} 
+ eEL^2q_y]\}.
\end{equation}
Here
$\Omega_c = eH/mc$ is the cyclotron frequency,
$f_N$ and  ${\cal N}_{q_{\perp}}$,
are the electron and phonon
distribution functions, respectively, 
$N = 0, 1, 2,...$ is the  Landau level (LL) index,
$q_{\perp} = \sqrt{q_x^2 +  q_y^2}$,
$s$ is the  velocity (its real part)
of leaky waves,
$\delta (q)$ is  the
 form-factor of LL's which at small their broadening 
$\Gamma$ can be assumed to be the Dirac delta function,
$V(q_{\perp})$ is the matrix element of the electron-phonon interaction, 
$|Q_{N,N^{\prime}}(L^2q^2_{\perp}/2)|^2  = 
|P_N^{N^{\prime} - N}(L^2q^2_{\perp}/2)|^2 \exp (- L^2q^2_{\perp}/2))$ 
is  determined by the overlap of the electron wave functions
corresponding to the initial and final states, and
$|P_N^{N^{\prime} - N}(L^2q^2_{\perp}/2)|^2$
is proportional to a Laguerre  polynomial.
The electron and phonon distribution functions
are assumed to be the Fermi and Planck functions, respectively,
with the temperature  $T$  (in energy units)
and   the Fermi energy $\zeta$ reckoned from the lowest LL.

The quantity  $I_{\Omega}(q_x,q_y)$ is proportional to
the incident microwave power.
It characterizes the effect
of microwave field on the in-plain electron motion.
Disregarding the influence of the microwave radiation polarization,
one can set~\cite{17,18}
$I_{\Omega}(q_x,q_y) = {\cal J}_{\Omega}L^2q^2_{\perp}$, where
${\cal J}_{\Omega} = ({\cal E}_{\Omega}/\tilde{{\cal E}}_{\Omega})^2$,
${\cal E}_{\Omega}$ is the microwave electric field amplitude,
which is assumed to be smaller than
some characteristic microwave field 
$\tilde{{\cal E}_{\Omega}}$,
and use the following formula~\cite{18}: 
\begin{equation}\label{eq2}
\tilde{{\cal E}}_{\Omega} = \overline{{\cal E}}_{\Omega}
\frac{|\Omega_c^2 - \Omega^2|L}
{e\Omega\sqrt{\Omega_c^2 +  \Omega^2}}.
\end{equation}
Here $ \overline{{\cal E}}_{\Omega} = \sqrt{2}m\Omega^2L/e$.
Equation~(1), which describes the effect of microwave field
by the inclusion of factor  $I_{\Omega}(q_x,q_y)$,
corresponds  to the single-photon absorption
of microwave radiation with the electron transitions  between different LL's.
The processes associated with the emission of microwave photons
involving   phonons in which
electrons do not transfer between LL's are  also be possible and can
provide some contribution to the photoconductivity at $\Omega \ll \Omega_c$.
However, the range  $\Omega \ll \Omega_c$
is not considered here.
Equations~(1) and (2) are valid
even in the vicinity of the cyclotron resonance $\Omega = \Omega_c$,
if ${\cal E}_{\Omega}/ \overline{{\cal E}_{\Omega}} < \Gamma/\Omega_c$, when 
the quantity $\tilde{{\cal E}_{\Omega}}$ is limited by the LL broadening.
In this limit one can estimate $\tilde{{\cal E}_{\Omega}} 
\simeq \sqrt{2}m\Omega\Gamma/e \overline{{\cal E}}_{\Omega}\Gamma/\Omega_c$.
In the case  when ${\cal E}_{\Omega}/ \overline{{\cal E}_{\Omega}} > 
\Gamma/\Omega_c$,
the dependence of the photon absorption
on the microwave electric field becomes more complex, particularly
 at the cyclotron resonance, because  multi-photon
processes (both real and virtual) become important.
In such a case,
the probability of the processes involving $n = 0, 1, 2,...$ real
photons are proportional to  
$I_{n\Omega}(q_x,q_y) = J_n^2(\sqrt{{\cal J}_{\Omega}}Lq_{\perp})$,
where $J_n(q)$ is the Bessel function~\cite{17,18}.

Using the  variables $q_{\perp}$ and $\Theta$ (instead of $q_x$ and $q_y$), 
so that 
$q_y = q_{\perp}\sin\Theta$, from Eq.~(1) we arrive at
$$
j_{ph}(E) = {\cal J}_{\Omega}\biggl(\frac{2\pi e}{\hbar^2sL}\biggr)  
\sum_{N,\Lambda>0}f_N(1 - f_{N+\Lambda})
$$
$$
\times\int_0^{2\pi}d\,\Theta \sin\Theta\int_0^{\infty} d\,q_{\perp}
q^2_{\perp}\exp(- L^2q^2_{\perp}/2)|V(q_{\perp})|^2
$$
$$
\times|P_N^{\Lambda}(L^2q^2_{\perp}/2)|^2 \{
{\cal N}_{q_{\perp}}
\delta[q_{\perp}  + q_{\Omega}^{\Lambda} + (eEL^2/\hbar s)q_{\perp}\sin\Theta]
$$
\begin{equation}\label{eq3}
+ ({\cal N}_{q_{\perp}} + 1)
\delta[q_{\perp}  - q_{\Omega}^{\Lambda} - (eEL^2/\hbar s)q_{\perp}\sin\Theta]\}
,
\end{equation}
where 
$q^{(\Lambda)}_\Omega{} = (\Omega - \Lambda\Omega_c)/s$, and $\Lambda > 0$.
At low electric fields, 
one can expand  the expression in the right-hand side 
of Eq.~(3) in powers of $(eEL^2/\hbar s)$. Upon integrating
over $\Theta$ we  present 
the dissipative dc photoconductivity $\sigma_{ph} = j_{ph}/E$ 
in the following form:
$$
\sigma_{ph} = {\cal J}_{\Omega}\biggl(\frac{\pi e^2L^3}{\hbar^3s^2}\biggr)  
\sum_{N,\Lambda>0}f_N(1 - f_{N+\Lambda})
$$
$$
\times\int_0^{\infty} d\,q_{\perp}
q^5_{\perp}\exp(- L^2q^2_{\perp}/2)
|V(q_{\perp})|^2|P_N^{\Lambda}(L^2q^2_{\perp}/2)|^2 
$$
\begin{equation}\label{eq4}
\times[
{\cal N}_{q_{\perp}}
\delta^{\prime}(q_{\perp}  + q_{\Omega}^{\Lambda})
- ({\cal N}_{q_{\perp}} + 1)
\delta^{\prime}(q_{\perp}  - q_{\Omega}^{\Lambda})]
,
\end{equation}
where $\delta^{\prime}(q)$ is the derivative of $\delta(q)$.
The inequality $(eEL^2/\hbar s) \ll 1$  implies that
the velocity of  electron Hall drift $v_H = cE/H \ll s$. 
\begin{figure}
\begin{center}
\includegraphics[width=7.8cm]{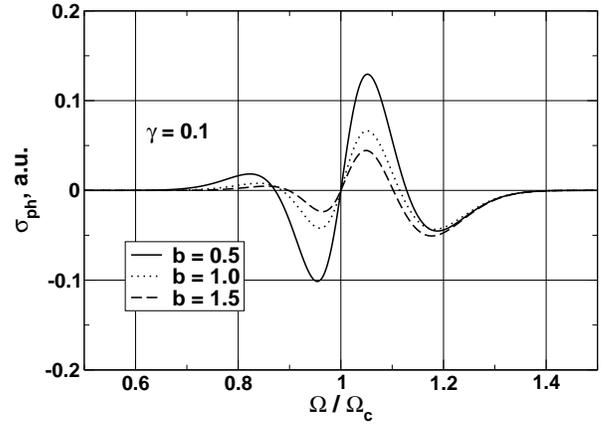}
\end{center}
\label{fig1}
\caption{Photoconductivity
vs microwave frequency 
for different $b = \hbar s/LT$
.
}
\end{figure}
At low temperatures $T \ll \hbar\Omega_c$
in some range of microwave frequencies around the $\Lambda$th resonance
$|\Omega - \Lambda\Omega_c| < \Omega_c$, where
only the  $\Lambda$th term in Eq.~(4) can be retained,
one can reduce Eq.~(4) 
to 
$$
\sigma_{ph} \simeq - {\cal J}_{\Omega}
\biggl(\frac{\pi e^2L^3}{\hbar^3s^2}\biggr)  
\sum_{N = N_m -\Lambda + 1}^{N_m}f_N(1 - f_{N+\Lambda})
$$
\begin{equation}\label{eq5}
\times \frac{d}{d\,q_{\perp}}\biggl[\frac{K_N^{\Lambda}(q_{\perp})}
{\exp(\hbar sq_{\perp}/T ) - 1}\biggr]
\biggr|_{q_{\perp} =  - q_{\Omega}^{\Lambda}}
\end{equation}
at $\Omega - \Lambda\Omega_c < 0$ (i.e., when $q_{\Omega}^{\Lambda} < 0$ ), and
$$
\sigma_{ph} \simeq {\cal J}_{\Omega}
\biggl(\frac{\pi e^2L^3}{\hbar^3s^2}\biggr)  
\sum_{N = N_m -\Lambda + 1}^{N_m}f_N(1 - f_{N+\Lambda})
$$
\begin{equation}\label{eq6}
\times \frac{d}{d\,q_{\perp}}
\biggl[\frac{K_N^{\Lambda}(q_{\perp})\exp(\hbar sq_{\perp}/T )}
{\exp(\hbar sq_{\perp}/T ) - 1}\biggr]
\biggr|_{q_{\perp} =   q_{\Omega}^{\Lambda}}
\end{equation}
at $\Omega - \Lambda\Omega_c > 0$ (when  $q_{\Omega}^{\Lambda} > 0$ ).
Here 
$K_N^{\Lambda}(q_{\perp}) = q^5_{\perp}\exp(- L^2q^2_{\perp}/2)
|V(q_{\perp})|^2|P_N^{\Lambda}(L^2q^2_{\perp}/2)|^2$
and
$N_m$ is the number of  filled LL's, i.e., $N_m\hbar\Omega_c < \zeta
 < (N_m + 1)\hbar\Omega_c$. 
For a large $N$, 
expressing  the Laguerre polynomials via the Bessel functions
and assuming that $|V(q_{\perp})|^2| \propto 1/q_{\perp}$, 
we have $K_N^{\Lambda}(q_{\perp}) \propto
q^4_{\perp}\exp(- L^2q^2_{\perp}/2)J_{\Lambda}^2(\sqrt{2N}Lq_{\perp})$.
Thus, $K_N^{\Lambda}(q_{\perp}) \propto
q^{4 + 2\Lambda}_{\perp}$
at $Lq_{\perp} < 1/\sqrt{2N}$,
and $K_N^{\Lambda}(q_{\perp}) \propto
q^{3}_{\perp}\exp(- L^2q^2_{\perp}/2)\cos^2[\sqrt{2N}Lq_{\perp} - (2\Lambda + 1)]$ at $Lq_{\perp} \gg 1/\sqrt{2N}$.

In the case of  resonance detuning 
$1/\sqrt{2N} < L|\Omega - \Lambda\Omega_c|/s \lesssim (LT/\hbar s)
\ll L\Omega_c/s$ (in the immediate vicinity of the resonance $|\sigma_{ph}|$
is very small),
from Eqs.~(4) and (5) we arrive at
\begin{equation}\label{eq7}
\sigma_{ph} \propto {\cal J}_{\Omega}\biggl(\frac{LT}{\hbar s}\biggr)
\frac{L(\Omega - \Lambda\Omega_c)}{s}.
\end{equation}
Here we have averaged  an  oscillatory factor in $K_N^{\Lambda}$
bearing in mind the finite broadening of the LL's.
In the range 
$(LT/\hbar s) < L|\Omega - \Lambda\Omega_c|/s \ll L\Omega_c/s$,
Eqs.~(5) and (6) yield
$$
\sigma_{ph} \propto  {\cal J}_{\Omega}
\frac{L^2(\Omega - \Lambda\Omega_c)^2}{s^2}
$$
$$
\times\biggl[ \frac{L^2(\Omega - \Lambda\Omega_c)^2}{s^2} 
- \biggl(\frac{\hbar s}{LT}\biggr)
\frac{L(\Omega - \Lambda\Omega_c)}{s} 
- 3\biggr]
$$
\begin{equation}\label{eq8}
\times\exp\biggl[\frac{\hbar(\Omega - \Lambda\Omega_c)}{T}\biggr]
\exp\biggl[-\frac{L^2(\Omega - \Lambda\Omega_c)^2}{2s^2}\biggr]
\end{equation}
at $\Omega - \Lambda\Omega_c < 0$, and

$$
\sigma_{ph} \propto  - {\cal J}_{\Omega}
\frac{L^2(\Omega - \Lambda\Omega_c)^2}{s^2}
$$
\begin{equation}\label{eq9}
\times\biggl[ \frac{L^2(\Omega - \Lambda\Omega_c)^2}{s^2} - 3\biggr]
\exp\biggl[-\frac{L^2(\Omega - \Lambda\Omega_c)^2}{2s^2}\biggr]
\end{equation}
at $\Omega - \Lambda\Omega_c > 0$.

The obtained formulas  describe  an oscillatory 
dependence of  the dissipative dc photoconductivity
associated with the photon-assisted interaction of electrons
with leaky interface phonons on the microwave radiation frequency 
and the magnetic field. 
Despite of a marked difference in the  dissipative 
dark conductivities 
associated with the 2D and 3D electron scattering on phonons, respectively,
(compare Refs.~\cite{16} and~\cite{14}),
the spectral dependence of the dissipative
dc microwave
photoconductivity calculated here is qualitatively
similar to that
obtained for the case of  the photon-assisted interaction of electrons
with 3D-acoustic phonons~\cite{14}.
Figure~1 shows the dissipative dc photoconductivity
as a function of microwave frequency calculated
using Eqs.~(5) and (6) for $N_m \gg 1$
and different values
of parameter $b =\hbar s/LT$, i.e., for different temperatures.
The frequency dependence of  ${\cal J}_{\Omega}$ was taken according
to Eq.~(2) with a proper modification at the immediate
vicinity of the cyclotron resonance:
${\cal J}_{\Omega} \propto \displaystyle
\frac{(\Omega_c^2 +  \Omega^2)}
{\Omega^2(\Omega_c+  \Omega)^2[(\Omega_c - \Omega)^2 + \Gamma^2]}$
with $\gamma = \Gamma/\hbar\Omega_c = 0.1$.
One can see that the dissipative dc photoconductivity
associated with the scattering processes under consideration
exhibits a pronounced minimum with $\sigma_{ph} < 0$
in the microwave frequency range
between the first (cyclotron) and the second resonances 
($\Omega_c < \Omega < 2\Omega_c$). At sufficiently strong radiation
with the frequency in this range,
the  value of the dissipative dc photoconductivity can exceed
the dark conductivity leading to ANC. 
The dissipative dc photoconductivity is negative also
in the ranges between higher resonances 
($\Lambda\Omega_c < \Omega < (\Lambda + 1)\Omega_c$ with $\Lambda > 1$). 
However, the amplitude
of the photoconductivity oscillations in these ranges is
markedly smaller due to a significant decrease in ${\cal J}_{\Omega}$
with increasing ratio $\Omega/\Omega_c$.
It is instructive that the ``phonon'' mechanisms
(considered above and in Ref.~\cite{14})
and the ``impurity'' mechanism~\cite{9,10,11}
result in quite different behavior of the microwave photoconductivity 
as a function of the resonance detuning, particularly in the vicinities
of the resonances.

The author is grateful to V.~A.~Volkov and V.~V.~Vyurkov 
for numerous discussions
and  R.~R.~Du and M.~A.~Zudov
for providing Refs.~\cite{4,15}.

\end{document}